\documentclass{article}
\usepackage[utf8]{inputenc}
\usepackage{jstyle}

\title{Solutions in Nonlinear Electrodynamics and their double copy regular black holes}
\author{Karapet Mkrtchyan,}
\author{Mantas Svazas}
\affiliation{Theoretical Physics Group, Blackett Laboratory, Imperial College London SW7 2AZ, U.K.}

\emailAdd{\mbox{ k.mkrtchyan@imperial.ac.uk, mantas.svazas20@imperial.ac.uk}}

\date{February 2022}

\abstract{We study solutions in non-linear electrodynamics (NED) and establish several general results.
We show, that the $SO(2)$ electric-magnetic duality symmetry is restrictive enough to allow for reconstruction of the NED Lagrangian from the spherically-symmetric electrostatic (Coulomb-like) solution --- although there are infinitely many different NED theories admitting a given solution, there exists a unique $SO(2)$ invariant one among them. 
We introduce a general algorithm for constructing new $SO(2)$ invariant NED theories in the conventional approach, where only a few examples are available. We also show how to derive the Lagrangian of the $SO(2)$ invariant theory admitting a given electrostatic solution.
We further show on a simple example that some NED theories may require sources (particles) of finite (non-zero) size. Such a non-zero size source not only regularizes the infinite energy of the point charge but also satisfies the condition of regularity, that the electric field is zero at the origin.
The latter condition was identified earlier as necessary and sufficient for the NED solution to generate a regular black hole via so-called double copy construction and is also satisfied by solitons.
We propose a large class of solitonic NED solutions that give rise to regular black holes via double copy construction and contain solutions of Maxwell and Born-Infeld as different limits.
This class of NED solutions acquires two new properties in the limit where the corresponding regular black hole's asymptotics becomes Minkowski: it gives rise to regular higher-spin black holes via generalization of double copy --- ``higher-copy'' construction, and for very short distances changes the sign of the force becoming repulsive/attractive for opposite/similar signs of charges.
}

\begin{document}

{\phantom{.}\vspace{-2.5cm}\\\flushright Imperial-TP-KM-2022-2\\
}

\maketitle

\section{Introduction}

Nonlinear electrodynamics (NED) is a source of inspiration since almost a century \cite{Rainich:1925,Born:1933qff,Born:1933pep,Born:1934gh,Heisenberg:1936nmg}, and despite of the large shift in our perspective on quantum field theory since then and lack of experimental support for non-linearities in pure electrodynamics,\footnote{See, however, \cite{Diego-Palazuelos:2022dsq}.} it remains one of the fascinating subjects that produces new ideas and methods in physics.

In Maxwell electrodynamics, the static spherically-symmetric solution (SSSS) is given by the Coulomb force of the point charge that grows infinitely in the proximity of the source. While this does not necessarily imply inconsistency of the theory (quantum electrodynamics is a well-defined theory with extremely precise experimental tests), one could consider a possibility of NED theory, where the field values are limited from above. This logic originally led Born and Infeld to develop their celebrated theory (BI theory) that has a maximal value of electric field \cite{Born:1934gh}. This theory was later shown to possess many remarkable features and reappeared in different contexts, most notably in String Theory \cite{Fradkin:1985qd}. One of the remarkable features of BI theory is that it is symmetric with respect to $SO(2)$ duality rotations \cite{Gaillard:1981rj,Bialynicki-Birula:1984daz,Gibbons:1995cv,Pasti:1995tn,Avetisyan:2021heg}, making it the second example historically in the literature after the free Maxwell theory. 

In more recent times, non-linear electrodynamics (NED) has been in the centre of attention of physicists for several interesting properties (for recent discussion see, e.g., \cite{Alam:2021ovb,Dehghani:2021fwb,Sorokin:2021tge,Babaei-Aghbolagh:2022uij}), in particular, for sourcing regular black holes (see, e.g., \cite{Ayon-Beato:1998hmi,Bronnikov:2000vy,Balart:2014cga}).

A completely different approach to constructing solutions for gravitational theories that may make use of NED emerged recently: the so-called double copy construction based on the remarkable property of gauge theory to provide formulas for observable quantities in gravity \cite{Bern:2008qj,Bern:2010ue,Johansson:2014zca} generalizing a phenomenon first observed in String Theory context \cite{Kawai:1985xq}. As opposed to other approaches where the NED was a source for gravity allowing for new solutions, in this approach, gauge theory (in particular, NED) is just an auxiliary tool that allows to compute quantities in gravity without actually being part of the gravitational theory  (for a recent review, see \cite{Adamo:2022dcm}). Here, Gravity is understood as a ``square'' of a gauge theory, with a concrete recipe on how to derive gravitational observables from the gauge theory ones (for a discussion on the Lagrangian formulation of the double copy, see \cite{Ferrero:2020vww} and references therein).
In the double copy approach, the Schwarzschild and Kerr black holes can be derived by double copy from solutions of Abelian Maxwell theory \cite{Monteiro:2014cda}. It is then natural to look for a deformation of the Maxwell theory that may provide for solutions whose double copy correspond to regular black holes \cite{Easson:2020esh}. The latter may in turn be a solution for some modification of Einstein gravity. A natural (and String-Theory motivated \cite{Tseytlin:1996it,Rocek:1997hi,Tseytlin:1999dj}) candidate is the exceptional NED --- Born-Infeld theory, which was examined in this light recently \cite{Pasarin:2020qoa}. There, the double copy metric based on the static SSSS of BI theory was shown to still contain curvature singularity. It was also understood that the condition for double copy metric to be regular is that the corresponding NED solution $\overrightarrow E(\overrightarrow r)$ has zero electric field at the origin: $\overrightarrow E(0)=0$. This condition is also required for everywhere continuous $\overrightarrow E(\overrightarrow r)$.
Everywhere continuous SSSS with nearly Coulomb behaviour at large distances will have to have maximal value $E=|\overrightarrow E|$ for the electric field at some distance $r_0\neq 0$ from origin, that will define a ``size'' for the corresponding ``particle''. Therefore, such theories have a scale, below which they are hardly distinguishable from Maxwell theory.

It is therefore clear from the results of \cite{Pasarin:2020qoa} that if a NED has a solitonic SSSS, its double copy will be a regular black hole. We will see in the following that solitons are not the only option.

Recently, in \cite{Avetisyan:2021heg}, a democratic formulation of NED was explicitly constructed that also manifests the $SO(2)$ duality-symmetry, when present. For the purposes of this work, we turn back to the conventional single-potential formulation, where the $SO(2)$ duality symmetry is not manifest. We will clarify some questions in this formulation and also construct double copy metrics for some NED solutions. This double copy construction of gravitational solutions is so far only known in the single potential formulation, where $S-$duality aspects are not transparent (see, however, \cite{Huang:2019cja,Alawadhi:2019urr}). We leave the intriguing question of adapting the double copy to democratic NED to a future work (there, one may need to consider a democratic approach to gravity \cite{Henneaux:2004jw} as well).

We organize this paper as follows. In Section \ref{Sol} we introduce generalities and some classes of solutions in NED, including the spherically-symmetric static solutions, that are central in this work.
In Section \ref{Algorithm} we present an algorithm to construct duality-symmetric NED theories in a standard formulation. In Section \ref{Examples} we discuss examples of NED theories and their spherically-symmetric static solutions with the properties that the electric field is bounded from above and goes to zero at the centre. In Section \ref{DScompletion} we show that for $SO(2)$ duality-symmetric NED theories the Lagrangian can be reconstructed from the data of spherically-symmetric static solution. We also provide an efficient derivation algorithm for such Lagrangians from the solution data, using the ansatz of Section \ref{Algorithm}. In Section \ref{DoubleCopy} we discuss the double copy construction of regular black holes from NED solutions. We conclude in Section \ref{Conclusion} with some outlook and open problems. In Appendix \ref{NoBirefringence} we rederive the no-birefringence condition and show that Born-Infeld theory is the only solution for it other than Maxwell, under mild assumptions. In Appendix \ref{Energy} we provide some details on how to compute the energy of the NED solutions.




\section{Solutions in NED}\label{Sol}

We consider non-linear electrodynamics (NED) described by an action \cite{Peres:1961zz,Plebanski:1970zz,Bialynicki-Birula:1984daz} of the form:
\begin{align}
    \mathcal{L}=\mathcal{L}(s,p)\,,\label{L}
\end{align}
where ($\mu, \nu,\dots=0,1,2,3$ are space-time indices, $\epsilon_{\mu\nu\alpha\beta}$ is the fully antisymmetric Levi-Civita tensor) 
\begin{align}
    s=\frac12\,F_{\mu\nu}\,F^{\mu\nu}\,,\quad p=\frac12\,F_{\mu\nu}\star F^{\mu\nu}\,,\\
    \star F_{\mu\nu}=\frac12\,\epsilon_{\mu\nu\alpha\beta}\,F^{\alpha\beta}\,,\quad F_{\mu\nu}=\partial_{\mu}\,A_{\nu}-\partial_{\nu}\,A_{\mu}\,.
\end{align}
The equations of motion for this theory are given as:
\begin{align}
\label{EquationsOfMotion}
    \partial^{\mu}(\beta(s,p)\, F_{\mu\nu}+\alpha(s,p)\,\star F_{\mu\nu})=0\,,
\end{align}
with
\begin{align}
    \beta(s,p)=\frac{\partial \mathcal{L}}{\partial s}\,,\quad
    \alpha(s,p)=-\frac{\partial \mathcal{L}}{\partial p}\,.
\end{align}
Electric and magnetic vector fields are given as ($i,j,\dots=1,2,3$ are space indices):
\begin{align}
    E_i=F_{0i}\,,\quad B_i=\frac12\,\epsilon_{ijk}\,F^{jk}\,.
\end{align}
Here we will study a general class of solutions, for which the vector-potential can be given in the following form:
\begin{align}
    A_\mu=\varphi(n\cdot x)\,k_\mu\,,\label{AmuKerr}
\end{align}
where $n_\mu$ and $k_\mu$ are given vectors.


\subsection{Wave solutions}

Even though in this paper we will be mainly interested in static spherically-symmetric solutions for NED, we remind here some facts about wave solutions.

An interesting observation is that arbitrary NED \eqref{L} admits wave solutions familiar from Maxwell electrodynamics\footnote{This property holds in general for NED of the type \eqref{L}, but is not always true for non-relativistic or higher-derivative NED's. We thank Iwo Bia\l ynicki-Birula for correspondence on this matter.} \cite{Peres:1960wfo,Plebanski:1970zz,Boillat:1970gw,Bialynicki-Birula:1984daz,BialynickiBirula:1992qj}. This rather remarkable fact can be shown by noticing that for constant $s$ and $p$, the NED equations \eqref{EquationsOfMotion} are equivalent to Maxwell equations:
\begin{align}
    \partial^\mu F_{\mu\nu}=0\,,
\end{align}
with familiar wave solutions. It is therefore immediate to conclude, that {\it those wave solutions of Maxwell theory, for which $s$ and $p$ are constant, are also solutions for any NED}. 

Such is, e.g., the solution \eqref{AmuKerr} for $k_\mu$ and $n_\mu$ satisfying
\begin{align}
    k_\mu\cdot n^\mu=0\,,\quad n_\mu\cdot n^\mu=0\,,
\end{align}
for which $s=0=p$. Such field configurations are known as ``null electromagnetic fields'' (see, e.g., \cite{Peres:1961zz}).
If we choose now $n_\mu=(1,n_i)$, where $n_i$ is a unit vector in a specified direction in space, the solution \eqref{AmuKerr} will describe a wave in the same direction, that solves any NED \eqref{L}.
However, as opposed to Maxwell theory, for NED the superposition of plane waves is not anymore a solution, unless they are in the same direction. Furthermore, the waves in NED experience vacuum birefringence, absent only in Born-Infeld electrodynamics \cite{Bialynicki-Birula:1984daz} (apart from cases that do not seem to have physical applications). We give more details on this in Appendix \ref{NoBirefringence}.

\subsection{Spherically Symmetric Static Solutions}

The main concern of this work will be the spherically symmetric electrostatic solutions given by \eqref{AmuKerr} and 
\begin{align}
    k_\mu=(1,x^i/r)\,,\qquad n_i=(0,x^i/r)\,,\qquad r=\sqrt{x_i\,x^i}\,,
\end{align}
therefore
\begin{align}
    A_{0}=\varphi(r)\,,\quad A_i=\varphi(r)\frac{x_i}{r}\,,\qquad F_{ij}=0\,,\quad F_{0i}=\varphi^\prime(r)\frac{x_i}{r}\,.
\end{align}
Electrostatic solutions in general satisfy the following equation in the vacuum:
\begin{align}
    B_i=0\,,\quad \partial^i(\beta(-E^2) E_i)=0\,,\quad \beta(s)=\beta(s,p)|_{p=0}\,,
\end{align}
while those with spherical symmetry satisfy ($r$ is the radial coordinate):
\begin{align}
    \frac{1}{r^2}\frac{\partial}{\partial r}(r^2\,\beta(-E^2)\,E(r))=0\,,\label{VacuumEq}
\end{align}
which is solved as:
\begin{align}
    -2\,\beta(-E^2)\,E(r)=\frac{Q}{r^2}\,,\label{Qbeta}
\end{align}
where $Q$ is a constant, corresponding to charge.

Now, the function $E(r)$ that defines the analogue of Coulomb field depends on the function $\beta(s)$, which in turn is defined by the Lagrangian $\mathcal{L}$.
Inversely, given the $E(r)$, one can look for $\beta(s)$, for which the equation \eqref{Qbeta} is satisfied, and then a Lagrangian \eqref{L} for which:
\begin{align}
    \beta(s)=\frac{\partial \mathcal{L}}{\partial s}|_{p=0}\,.
\end{align}
The latter identification is not unique: there are many different theories of non-linear electrodynamics that have the same electrostatic solution. 

In the Maxwell (or Born-Infeld) electrodynamics, the spherically symmetric solution determines the vacuum field everywhere, except for one point, where the point charge source should be introduced.
One alternative to this scenario is solitonic solutions that do not require sources, which we will discuss below on examples. 
We will demonstrate on a simple example that there is one more possibility: the vacuum solution can be determined everywhere except for a ball of a finite radius, requiring a charged source of finite size. 

We also show how to compute the energy of the electrostatic spherically-symmetric solutions in Appendix \ref{Energy}.

\section{An algorithm of constructing duality-symmetric NEDs}\label{Algorithm}

In this section, we describe a method to obtain a duality symmetric Lagrangian of non-linear electrodynamics (see also \cite{Svazas:2021ltf}) in the conventional single-potential form.     
We first introduce $u,v$ variables as follows (see, e.g., \cite{Bandos:2020hgy}):
\begin{align}
    &u = s + \sqrt{s^2 + p ^2}\geq 0\,, &v = - s + \sqrt{s^2 + p^2 } \geq 0\,.
\end{align}
The Lagrangian of the form \eqref{L} can be now rewritten as $\mathcal{L}(u,v)$ and is duality symmetric
if it satisfies the duality-symmetry constraint (see, e.g., \cite{Gibbons:1995cv}),
\begin{align}
    \frac{\partial \mathcal{L}}{\partial u}\, \frac{\partial \mathcal{L}}{\partial v}=-1\,.\label{CHE}
\end{align}
Only a handful of explicit analytic solutions to this equation are known so far (see, e.g., \cite{Hatsuda:1999ys,Gaillard:1997zr,Ivanov:2003uj,Bandos:2020hgy}) apart from Maxwell and Born-Infeld. The democratic formulation of duality-symmetric theories with explicit Lagrangian featuring an arbitrary function of one variable is given in \cite{Avetisyan:2021heg,Avetisyan:2022zza}. Translation of these democratic theories into the conventional approach with single vector-potential still requires solving the equation \eqref{CHE}.

We choose an ansatz for the Lagrangian $\mathcal{L}(u,v)$ in the following form
\begin{align}
    \mathcal{L}(u,v) = 2\sqrt{v(\rho - u)} - f(\rho)\,,\label{Luv}
\end{align}
with some function $\rho(u,v)$ and $f(\rho)$ that will be specified later for concrete examples.
The advantage of the ansatz \eqref{Luv} is that the duality-symmetry \eqref{CHE}
can be achieved when:
\begin{align}
    f'(\rho) \equiv \frac{\partial f}{\partial \rho} = \sqrt{\frac{v}{\rho - u}}\,.\label{Eq:rho}
\end{align}
The latter can be understood as an equation that should be solved to find $\rho(u,v)$. We therefore can choose $f(\rho)$ arbitrarily and find corresponding duality-symmetric theory by solving \eqref{Eq:rho}. This gives us a map from the space of functions of one variable to the duality-symmetric theories of electrodynamics --- the space of solutions of \eqref{CHE}, specified by a function of one variable. 

When computing the equations of motion in the form of (\ref{EquationsOfMotion}), the $\alpha$ and $\beta$ are expressed as:
\begin{align}
    \alpha &= \bigg[\sqrt{\frac{v}{\rho - u}} + \sqrt{\frac{\rho - u}{v}}\bigg]\frac{\text{sign}(p)\sqrt{uv}}{u + v} = \bigg[f'(\rho) + \frac{1}{f'(\rho)}\bigg]\frac{\text{sign}(p)\sqrt{uv}}{u + v}\,, \\[5pt]
    \beta &=  \bigg[u\sqrt{\frac{v}{\rho - u}} - v\sqrt{\frac{\rho - u}{v}}\bigg]\frac{1}{u + v} = \bigg[uf'(\rho) - \frac{v}{f'(\rho)}\bigg]\frac{1}{u + v}\,.
\end{align}
Several examples of new duality-symmetric theories were derived using this approach in \cite{Svazas:2021ltf}.

\section{Regular electric field: Examples}\label{Examples}

A spherically symmetric solution being regular everywhere would require the electric vector field to be zero at the center.
This implies that everywhere continuous solution for $E_r(r)$ that has Coulomb asymptotics $\sim 1/r$ at $r\to \infty$ has to have an extremum at some finite radius $r_0$ and go to zero at $r=0$. This requires changing the Coulomb solution in a manner, somewhat similar to the Planck's resolution of the ultraviolet catastrophe in black body radiation. 

We will discuss two examples that demonstrate physically distinct properties as compared to Maxwell theory.

\subsection{Arcsinh electrodynamics}

We start from the following Lagrangian:\footnote{A similar example (with $\arcsin$ instead of $\text{arcsinh}$) is studied in \cite{Kruglov:2016ezw}, which, however, does not share the specific feature of the example studied here: the spherically symmetric solution requiring a non-zero size source.}
\begin{align}
    \mathcal{L}=-\frac12\, T\,\text{arcsinh}\Big(\frac{s}{T}\Big)\,,\label{ArcsinhLag}
\end{align}
whose large $T$ (or small field) limit coincides with Maxwell theory
\begin{align}
  \lim_{T\to\infty}\mathcal{L}=-\frac12\,s\,.
\end{align}
The dimensionful parameter $T$ has somewhat similar role to the tension in String Theory.
We get:
\begin{align}
    \beta(s)=-\frac12\,\frac{T}{\sqrt{T^2+s^2}}\,,
\end{align}
and the corresponding spherically symmetric solution satisfies the following equation:
\begin{align}
    \frac{T\,E}{\sqrt{T^2+E^4}}=\frac{Q}{r^2}\,.
\end{align}
For simplicity, we introduce dimensionless version of the electric field, $\bar E=E/\sqrt{T}$, and we also denote $r_0^2=\sqrt{\frac{2\,Q^2}{T}}$. Then, we get:
\begin{align}
    \sqrt{\frac{2\,\bar E^2}{1+\bar E^4}}=\frac{r_0^2}{r^2}\,,
\end{align}
with a solution (we introduce $\bar r=r/r_0$):
\begin{align}
    \bar E(\bar r)=\text{sign}(Q)\sqrt{\bar r^4 \pm \sqrt{\bar r^8-1}}
\end{align}
One of the solutions is growing at infinity, therefore we can discard it as unphysical (though it is interesting to notice that there is a second solution as compared to Maxwell theory, despite the fact that the theory under consideration is a deformation of Maxwell theory by higher order terms in field strength: the second solution is a non-perturbative effect).
We concentrate instead on the physical solution (we take $Q>0$ for simplicity here):
\begin{align}
    \bar E(\bar r)=\sqrt{\bar r^4-\sqrt{\bar r^8-1}}\,,
\end{align}
or,
\begin{align}
    E(r)=\sqrt{T}\,\sqrt{\Big(\frac{r}{r_0}\Big)^4-\sqrt{\Big(\frac{r}{r_0}\Big)^8-1}}\,.
    \label{Er1}
\end{align}
Note, that for large $\bar r>>1$ (or $r>>r_0$), the field strength is close to Coulomb one:
\begin{align}
    \bar E(r)=\frac{r_0^2}{\sqrt{2}\,r^2}+\frac{r_0^{10}}{8\sqrt{2}\,r^{10}}+O\Big((r_0/r)^{18}\Big)\,,
\end{align}
or,
\begin{align}
    E(r)=\frac{Q}{r^2}+\frac{1}{2\,T^2}\Big(\frac{Q}{r^2}\Big)^5+O\Big(\frac{Q^9}{T^4\,r^{18}}\Big)\,.
\end{align}
The latter equation contains modification to Coulomb law which is tiny for
\begin{align}
    r>>r_0=\sqrt[4]{2\,Q^2/T}\,,\label{r0}
\end{align}
but becomes significant for $r$ values comparable to $r_0$.

The remarkable feature of the solution \eqref{Er1} is that $E(r)$ is not real for $r<r_0$ (note, that this is true also for the second solution that we discarded). In order for $E(r)$ to be real everywhere, one needs to introduce a source in the equation \eqref{VacuumEq} for $r<r_0$.\footnote{This can indicate that the theory is incomplete: it cannot define the field almost everywhere except for point charged sources as it is customary for field theory. It is not clear how to define a theory for the sources of this field --- it cannot be a regular field theory. We thank Arkady Tseytlin for helpful discussions on this.} We are talking about a finite-size charged source, which has internal structure, given by a charge density function $\rho(r)$. The equation with source,
\begin{align}
    -2\,\frac1{r^2}\frac{\partial}{\partial r}(r^2\,\beta(E^2)\,E(r))=4\,\pi\,\rho(r)\,,\label{SourceEq}
\end{align}
should be considered instead of the vacuum equation \eqref{VacuumEq} for $r<r_0$.

Note, that the internal structure of the finite-size source is not given by the theory: different choices of the function $\rho(r)$ in the region $r\leq r_0$ could work. If we take, e.g., $\rho\sim r^n$ at $r\leq r_0$, we can see that for any integer $n\geq -1$ we can find a real solution for $E(r)$. For $\rho\sim 1/r$, we get that the $E(r)=const$ for $r\leq r_0$. This situation is somewhat similar to Born-Infeld (BI) theory, except that for BI theory $r_0=0$. The constant $E(r)=const\neq 0$ around $r=0$ means that the field is not defined at one point $r=0$ (remember that $E$ is a vector and though it has the same absolute value, but has different directions depending on how we take the limit $r\to 0$). Therefore, the case $\rho \sim 1/r$ does not have everywhere continuous solution.
Instead, sufficiently reasonable choices are $\rho\sim r^n$ with $n\geq 0$, for which we get non-singular solutions that can be glued to the vacuum solution at $r=r_0$, and we also get $E(0)=0$.

A simple choice is constant charge density ball of radius $r_0$ ($\bar \rho$ is a constant):
\begin{align}
\rho(r)=\begin{cases}
        \bar \rho\, & r<r_0\,,\\
        0\, & r\geq r_0\,.
        \end{cases}
\end{align}
The internal solution in this case has to solve the following equation:
\begin{align}
    \sqrt{\frac{2}{1+\bar E^4}}\bar E=\sigma\,r+\frac{\delta}{r^2}\,,
\end{align}
where $\sigma=\frac{4\,\pi}{3}\bar \rho\,\sqrt{\frac{2}{T}}$ and $\delta$ is a constant of integration, analogue of $Q$ for vacuum solution. It is straightforward to see that there is a regular solution around $r=0$ only if $\delta=0$. Therefore, we get:
\begin{align}
    \bar E(r)=\text{sign}(\sigma)\sqrt{\frac{1}{\sigma^2\,r^2}-\sqrt{\frac{1}{\sigma^4\,r^4}-1}}\,,
\end{align}
In order for this solution to be glued to \eqref{Er1} at $r=r_0$ and give a continuous function for $E(r)$, we need $\text{sign}(\sigma)=\text{sign}(Q)$ and $\frac{1}{\sigma^2}=r_0^2$. The latter implies: 
\begin{align}
Q=\frac{4\pi}{3}\bar\rho\, r_0^3=4\,\pi\,\int_0^{r_0} \rho(r)\,r^2\,dr\,,
\end{align}
as expected. Then, we get (taking $Q>0$ for simplicity):
\begin{align}
    \bar E(\bar r)=\begin{cases}
    \sqrt{\frac{1}{\bar r^2}-\sqrt{\frac{1}{\bar r^4}-1}}\qquad &  0\leq \bar r\leq 1\,,\\
    \sqrt{\bar r^4-\sqrt{\bar r^8-1}}\qquad & \bar r\geq 1\,,
    \end{cases}
\end{align}
or,
\begin{align}
    E(r)=
    \begin{cases}
    \sqrt{T}\,\sqrt{\frac{r_0^2}{r^2}-\sqrt{\frac{r_0^4}{r^4}-1}}\qquad &  0\leq r\leq r_0\,,\\
    \sqrt{T}\,\sqrt{ \frac{r^4}{r_0^4}-\sqrt{\frac{r^8}{r_0^8}-1}}\qquad & r\geq r_0\,,
    \end{cases}\label{Ex1}
\end{align}
The maximal value of the field is reached at $r=r_0$ and is $E_{\text{max}}=\sqrt{T}$. Thus, the value of $T$ is related to maximal energy density of the electromagnetic field.

\begin{figure}[h]
\begin{center}
\includegraphics[width=14cm]{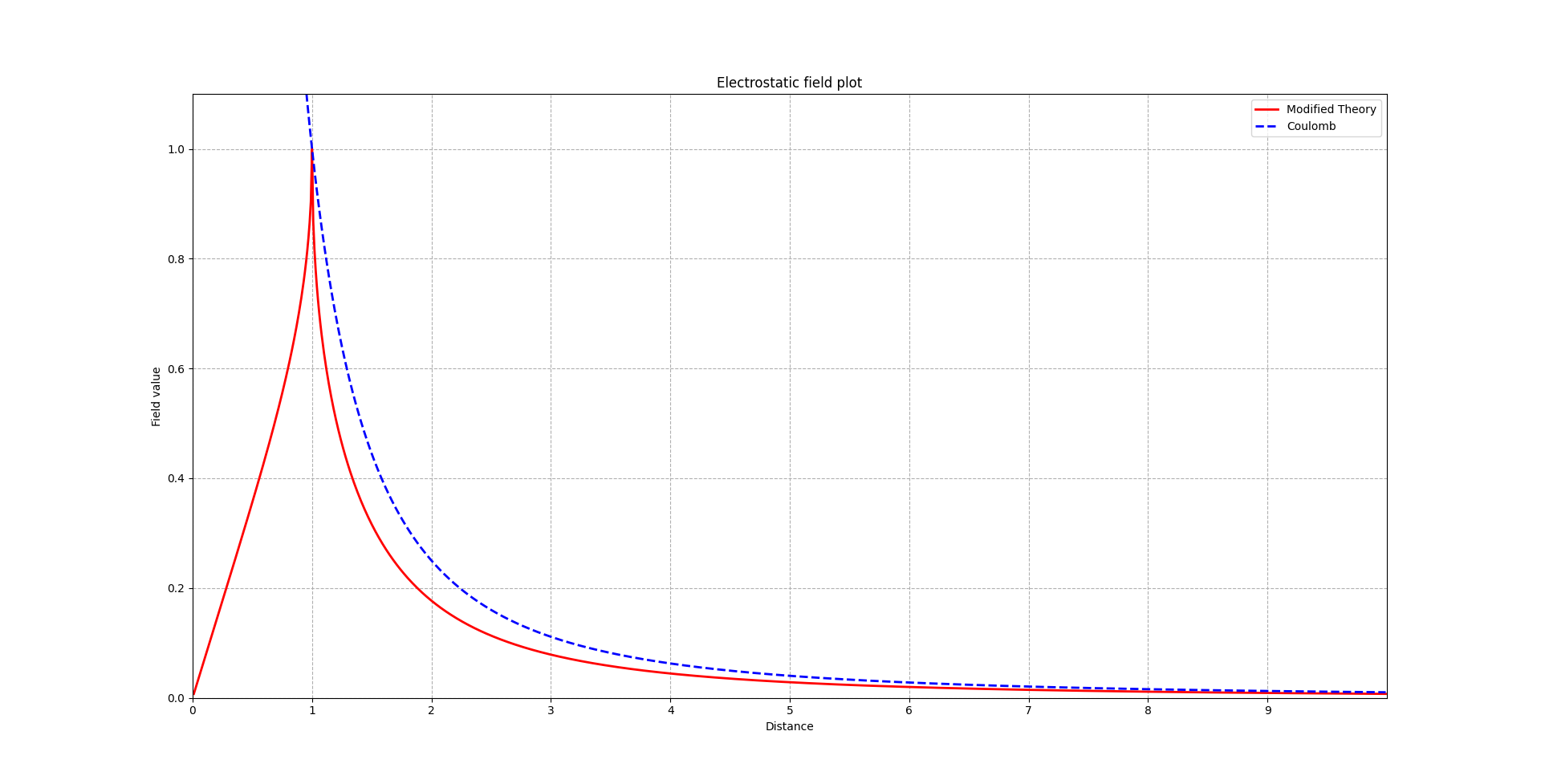}\caption{The solution \eqref{Ex1} compared to Coulomb solution.}
\end{center}
\end{figure}

The magnetic solutions can be found by first finding an electric-magnetic dual theory and then solving it in a similar manner. The dual theory is defined by introducing the dual curvature via the equation:
\begin{align}
   F^d_{\mu\nu}=\epsilon_{\mu\nu\lambda\rho}\, \frac{\partial \mathcal{L}}{\partial F_{\lambda\rho}}=-\frac{1}{\sqrt{1+s^2/T^2}}\,\star F_{\mu\nu}\,,
\end{align}
and then finding the $F$ as a function of $F^d$ (here $s_d=F^d_{\mu\nu}F^{d\,\mu\nu}/2$):
\begin{align}
    F_{\mu\nu}=-\sqrt{\frac{2}{1+\sqrt{1-4\,s_d^2/T^2}}}\,\star F^d_{\mu\nu}\,.
\end{align}
Then the Bianchi identity $dF=0$ becomes the dynamical equation for the dual field.
For the dual theory we get
\begin{align}
    \beta(s_d)=-\frac12\,\sqrt{\frac{2}{1+\sqrt{1-4\,s_d^2/T^2}}}\,,
\end{align}
which implies the following solution:
\begin{align}
    E^d(r)=\frac{Q_d\,r^2}{\sqrt{r^8+(r_0^{d})^8}}\,,\qquad r_0^d=\sqrt[4]{\frac{Q_d^2}{T}}\,.
\end{align}
Interestingly enough, this solution is solitonic: it does not require sources at all, but still has a characteristic size $r_0^d$.
The Lagrangian for the dual theory can be given as:
\begin{align}
    \mathcal{L}^d=\frac{1}{2}\,T\, \tanh^{-1}\left(\sqrt{\frac{1}{2} \left(1-\sqrt{1-\frac{4 s^2}{T^2}}\right)}\right)-T\,\sqrt{\frac{1}{2} \left(1-\sqrt{1-\frac{4 s^2}{T^2}}\right)}\,.
\end{align}
This Lagrangian, again, goes to Maxwell one, $\mathcal{L}\sim s/2$ (up to a sign factor), for $T\to \infty$. For this dual theory, the electric solution is solitonic while the magnetic one requires finite-size sources. It would be interesting to study the solution of this theory for generic dyonic charges,\footnote{We expect that depending on the ratio of electric and magnetic charges the solution will be either solitonic or require finite-size source. Then, for a specific ratio it may require a point source, which is puzzling from the perspective of the double copy black hole singularity.} but we will not attempt it here.

\subsection{Spherically symmetric static soliton}

\begin{figure}[h]
\begin{center}
    \includegraphics[width=14cm]{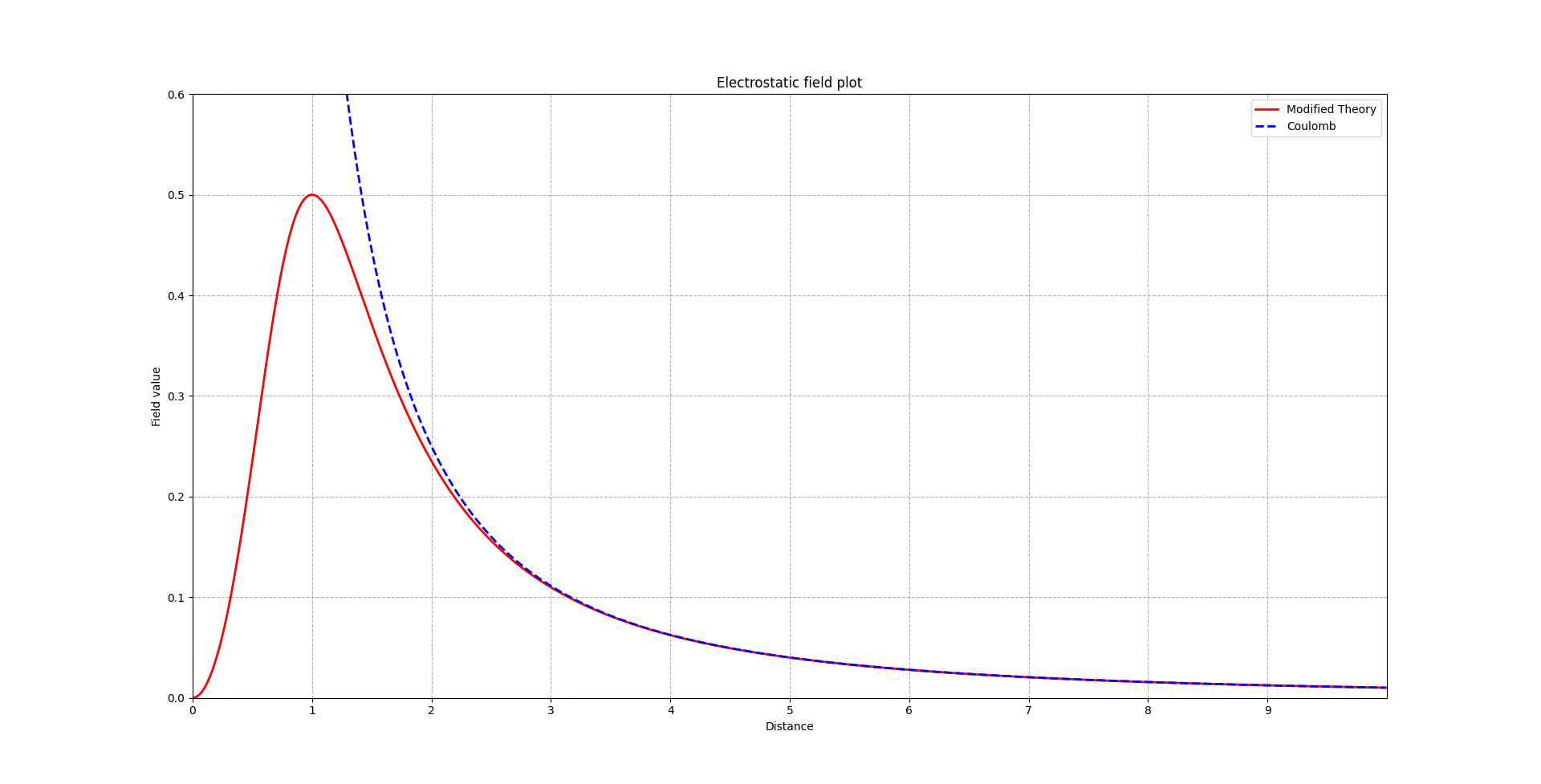}\caption{The solution \eqref{Ex2} compared to Coulomb solution.}
\label{Fig2}
\end{center}
\end{figure}

Here we will discuss another example of a solitonic vacuum solution that is regular and real everywhere. First one needs to choose a solution that modifies Coulomb only at short distances. Here, we will choose it to be of the form (see Figure \ref{Fig2}):
\begin{align}
    E(r)=\frac{Q\,r^2}{r^4+r_0^4}=\frac{Q}{r_0^2}\,\frac{1}{\bar r^2+\frac{1}{\bar r^2}}=E_0\,\frac{1}{\bar r^2+\frac{1}{\bar r^2}}\,,\label{Ex2}
\end{align}
where $\bar r=r/r_0$ and $E_0=Q/r_0^2$. Again, the difference compared to Coulomb law is small for $r>>r_0$.

The non-linear electrodynamics Lagrangian allowing for such a solution is not unique. One such Lagrangian can be given in the following form:
\begin{align}
\label{Example 2}
    \mathcal{L}=2\,T\, (\sqrt{1-s/T}-1)-2\,T\, \log \left(\frac{1+\sqrt{1-s/T}}2\right)\,.
\end{align}
We chose this Lagrangian among many possible ones by the condition that it is a function $\mathcal{L}(s)$ of $s$ only (no $p$ dependence) and $T\to \infty$ limit gives Maxwell Lagrangian $\mathcal{L}=-s/2$.

The theory given by \eqref{Example 2} admits solitonic electrostatic solution \eqref{Ex2} with finite size and energy. This is not the only Lagrangian that admits this solution though. We will derive later another Lagrangian with $SO(2)$ duality symmetry that admits the solution \eqref{Ex2}.

\subsection{The role of $\beta(s)$ in regular solutions}

An interesting general property of a regular (solitonic) vacuum solution with spherical symmetry reproducing Cuolomb behaviour at large distances is that it has to have an extremum at some point, where:
\begin{align}
    \frac{\partial E(r)}{\partial r}|_{r=r_0}=0\label{E'r0}
\end{align}
Note, that the equation \eqref{Qbeta} then implies:
\begin{align}
    \frac{\partial \beta}{\partial r}|_{r=r_0}\,E_0=\frac{Q}{r_0^3}\,.
\end{align}
Given that $E_0=E(r_0)$ is a finite number, we conclude that $\partial \beta/\partial r$ is finite and non-zero at $r=r_0$. On the other hand, we have:
\begin{align}
    \frac{\partial \beta}{\partial r}=\frac{\partial\beta}{\partial E}\,\frac{\partial E}{\partial r}\,.
\end{align}
Given \eqref{E'r0},
in order for $\partial \beta/\partial r$ to be non-zero, $\partial \beta/\partial E$ should diverge at $E=E_0$. The physical interpretation of $\beta$ is that it describes the permittivity of the medium. In the vacuum solution described here, the field value is restricted to $E < E_0$, thus a ``medium'' with $E > E_0$ cannot be the vacuum, so the divergence of $\partial \beta/\partial E$ could be interpreted as change of medium at the boundary of regions $E<E_0$ and $E>E_0$.\footnote{When studying electrodynamics in some medium, one encounters different non-linearities. While the theories studied here had been considered as fundamental field theories, one can take them also as effective theories of electrodynamics in a medium. Then, by appropriate choice of non-linear medium one can reproduce quasistable states described by the same equations as the solutions discussed here, but of macroscopic size. If such states can be realized in Nature, they could, e.g., provide an explanation of a ball lightning phenomenon.}


\section{Duality-symmetric Lagrangian for given spherically symmetric solution}\label{DScompletion}

For given spherically symmetric solution, there can exist many theories that admit this solution. Given solution depends on the function $\beta(s)=\frac{\partial \mathcal{L}}{\partial s}|_{p=0}$ which can be reconstructed from $E(r)$ almost uniquely \cite{Peres:1960wfo}, but not the Lagrangian. We will show below that the data about spherically symmetric solution (more precisely, $\beta(s,0)$), together with the $SO(2)$ duality-symmetry condition, fixes the Lagrangian completely.

\subsection{Duality-symmetric completion: proof of existence and uniqueness}

We would like to show that there is at most one $SO(2)$ duality-symmetric Lagrangian for given $\beta(s)$.
We will assume that the Lagrangian can be Taylor expanded in powers of $p$:
\begin{align}
    \mathcal{L}(s,p)=L_0(s)+p\,L_1(s)+\frac{p^2}{2}\,L_2(s)+\dots\,.\label{Lp}
\end{align}
Then,
\begin{align}
    \beta(s)= L_0^\prime(s)\,,
\end{align}
defines the electrostatic solutions. The duality-symmetry equation \cite{Bialynicki-Birula:1984daz} that the Lagrangian $\mathcal{L}(s,p)$ should satisfy, is 
\begin{align}
    \mathcal{L}_s^2-\frac{2\,s}{p}\mathcal{L}_s\,\mathcal{L}_p-\mathcal{L}_p^2=1\,,\label{DS}
\end{align}
where
\begin{align}
    \mathcal{L}_s=\frac{\partial \mathcal{L}}{\partial s}\equiv \beta(s,p)\,,\quad \mathcal{L}_p=\frac{\partial \mathcal{L}}{\partial p}\equiv -\alpha(s,p)\,.
\end{align}
As a side note, real solutions $\mathcal{L}_p$ of the quadratic equation \eqref{DS} require $\mathcal{L}_s^2\geq \frac{p^2}{s^2+p^2}$.
In terms of \eqref{Lp}, the equation \eqref{DS} can be rewritten as a series of equations for $L_n(s)$. One immediate conclusion is that for solutions of \eqref{DS}, $L_{2k+1}=0$ \footnote{This implies that all of the duality-symmetric theories are given through parity even Lagrangians $\mathcal{L}(s,p)$.}
. Then, the expansion \eqref{Lp} can be rewritten as:
\begin{align}
    \mathcal{L}(s,p)=\sum_{n=0}^{\infty} \frac{p^{2n}}{(2n)!}\,L_{2n}(s)\,,\label{Lsp}
\end{align}
and the equation \eqref{DS} solved as:\footnote{Naively, the equation \eqref{DS} could admit two solutions, but the analyticity condition in $p$ for the $\mathcal{L}(s,p)$ does not leave room for the second one. This condition is justified as we are discussing theories with non-singular $\beta$ at $p=0$.}
\begin{align}
    L_{2l+2}=\frac{2l+1}{2\,s}\Big(L^\prime_{2l}-\frac{1}{L^\prime_0}\Big)+\sum_{k=1}^{l}\frac{(2l+1)!}{(2k)!\,(2l-2k)!}\,\frac{1}{2\,s\,L_0^\prime}\Big[L_{2k}^\prime\,L_{2(l-k)}^\prime-\frac{2\,s}{2(l-k)+1}L_{2k}^\prime\,L_{2(l-k)+2}\nonumber\\-\frac{2k}{2(l-k)+1}L_{2k}\,L_{2(l-k)+2}\Big]\,,\qquad\qquad\label{L2l+2}
\end{align}
expressing all the $L_{2n}$ for $n>0$ via $L_0^\prime$.
We can therefore conclude that duality symmetry, represented by the equation \eqref{DS}, uniquely fixes the Lagrangian up to an irrelevant additive constant once the function $L_0^\prime(s)$ is given (note, that the function $L_0(s)$ itself is not needed to find all $L_{2n}(s)$, only its derivative enters the equations). On the other hand, $L_0^\prime(s)=\beta(s)$ is exactly the function entering the electrostatic equation \eqref{VacuumEq}. Thus, given an electrostatic solution, one has the $\beta(s)$ and can find the unique duality-symmetric Lagrangian corresponding to that electrostatic solution. 

Even though we gave a proof for existence of the duality symmetric theory with given solution, the equations \eqref{Lsp} and \eqref{L2l+2} are not easy to evaluate in the concrete examples to get a compact Lagrangian (we ignore the convergence issues that might obstruct the existence of an analytic solution). We discuss in the next section another method of deriving duality-symmetric Lagrangians from given $\beta(s)$.

\subsection{Duality-symmetric completion: an explicit derivation algorithm}

Now we would like to show, that for a known spherically-symmetric vacuum solution we can explicitly reconstruct a duality-symmetric Lagrangian admitting such a solution, using the ansatz \eqref{Luv}. The corresponding vacuum solution can be computed from equations by setting $u \to 0$ and $v \to -2s$. We denote as $\rho(s)\equiv \rho(u=0,v=-2s)$.
Then, the function $\beta$ relates to $\rho(s)$ and $f(\rho(s))$ as
\begin{align}
    \beta(s) = -\sqrt{\frac{-\rho(s)}{2s}} =- \frac{1}{f'(\rho)}\,,\label{beta f}
\end{align}
or
\begin{align}
    \rho(s) = -2\,s\,\beta^2(s)\,.\label{rho s}
\end{align}
For known $\beta(s)$, one solves \eqref{rho s} for $s(\rho)$, then uses \eqref{beta f},
\begin{align}
    \beta(s(\rho))=-\frac{1}{f'(\rho)}\,,
\end{align}
to find $f(\rho)$, and finally the duality-symmetric Lagrangian by solving \eqref{Eq:rho} for $\rho(u,v)$. 

We use the procedure outlined above for the example \eqref{Ex2}.
We get in this case:
\begin{align}
    f(\rho)=-4 T \log (1-\rho/2 T)
\end{align}
and the duality symmetric Lagrangian
\begin{align}
    \mathcal{L}(\bar u,\bar v)=4\, T\,\left( \sqrt{1+\bar v(1-\bar u)}+ \log \left(\frac{2 \left(\sqrt{1+\bar v(1-\bar u)}-1\right)}{\bar v}\right)-1\right)\,,
\end{align}
where $\bar u=u/(2T),\, \bar v=v/(2T)$ are the dimensionless variables. 

The $T\to \infty$ (or small field) limit of this theory is not Maxwell, but ModMax \cite{Bandos:2020jsw,Kosyakov:2020wxv} with $\gamma =-\log 2$:
\begin{align}
    \mathcal{L}=-\frac54\,s-\frac34\,\sqrt{s^2+p^2}\,.
\end{align}

Note, that the procedures described above may sometimes require solving complicated algebraic equations for which there are either many solutions or no solution can be written in analytic form. On the other hand, some of these algebraic complications might be related to the choice of the variables. In particular, a manifestly duality-symmetric description of \cite{Avetisyan:2021heg} may have much simpler expressions for a given case than the standard formulation with a single vector-potential.


\section{Double copy metrics for regular black holes}\label{DoubleCopy}

As discussed above, we get two possible scenarios where the non-linear electrodynamics allows for non-singular sphericaly-symmetric static solution $E(r)$ with $E(0)=0$.
Interestingly enough, this is exactly the condition \cite{Pasarin:2020qoa} that allows for a double copy construction of a regular black hole metric.
This can be formulated as: removing the singularity in electrostatic solution of the non-linear electrodynamics removes also the singularity in the corresponding double copy black hole metric.
In the gravitational solution, the singularity is presenting itself in the infinities of curvature invariants, while in NED the corresponding singularities are represented by point-charges and can be there even when the corresponding gauge invariants are limited from above. 
While the physical consistency of the scenario with finite-size sources (e.g., of the theory given by \eqref{ArcsinhLag}) is questionable, the soliton solutions have no consistency problems and are natural candidates for the double copy construction of regular black holes. 

We have established previously, that the NED theory can be reconstructed from a given profile of the electrostatic solution, therefore almost any function $E(r)$ can be a solution of some NED. We have also shown that there always exists a NED Lagrangian (even one with $SO(2)$ symmetry) for a given solution $E(r)$. We do not need explicit NED Lagrangian in the context of double copy: we treat the NED as an auxiliary construct, not involved in the gravitational theory. Instead, we need to make sure that the solution is close enough to Coulomb at large distances (so that the double copy metric is close enough to Schwarzschild) and that it generates a double copy metric of a regular black hole with Minkowski asymptotics.

\subsection{A general class of NED solitons}

A large class of NED solutions can be given by the following electrostatic configuration:
\begin{align}
    E(r)=\frac{Q\,r^n}{(r^{\frac{n+2}{m}}+r_0^{\frac{n+2}{m}})^m}\,,\label{GCEr}
\end{align}
which reproduces Coulomb for $r_0=0$ and Born-Infeld for $n=0,\,m=1/2$. For any $r_0,m,n>0$ we get a regular (soliton) electric field and can derive from it a regular double copy metric as follows.

We first compute the electric potential as:
\begin{align}
    \phi(r)=\int_r^{\infty} E(x)\,dx=\frac{Q}{r}\, _2F_1\left(m,\frac{m}{n+2};\frac{m}{n+2}+1;-\left(\frac{r_0}{r}\right)^{\frac{n+2}{m}}\right)\label{phi exp inf}
\end{align}
which can be rewritten as:
\begin{align}
    \phi(r)=\frac{Q}{r_0}\left[\frac{\Gamma \left(\frac{m (n+1)}{n+2}\right) \Gamma \left(\frac{m}{n+2}+1\right)}{\Gamma (m)}-\frac{1}{(n+1)}\left(\frac{r}{r_0}\right)^{n+1} {}_2F_1\left(m,\frac{m (n+1)}{n+2};\frac{m (n+1)}{n+2}+1;-\left(\frac{r}{r_0}\right)^{\frac{n+2}{m}}\right)\right]\label{phi exp 0}
\end{align}
It is now obvious that by choosing $n$ appropriately, e.g. integer, we can have a Taylor expansion of $\phi(r)$ around zero,
\begin{align}
    \phi(r)=c_0+c_1\,r+c_2\,r^2+c_3\,r^3+\dots\,,
\end{align}
with
\begin{align}
    c_1=c_2=\dots =c_n=0\,,
\end{align}
but $c_0\neq 0$.
As explained in \cite{Pasarin:2020qoa}, the $c_0$ can be removed by a large gauge transformation, which will change the asymptotics of the metric at infinity. Therefore, regular black hole metrics following from such a construction cannot be asymptotically Minkowski. 

When using \eqref{phi exp inf} in the double copy metric:
\begin{align}
    g_{\mu\nu}=\eta_{\mu\nu}+\phi(r)\,k_{\mu}\,k_{\nu}\,,\qquad k_{\mu}=(1,x_i/r)\,,\label{KS}
\end{align}
the comparison with Schwarzschild at large distances implies the replacement $Q\to 2\,G\,M/c^2=r_s$, where $G$ is the Newton constant, $M$ is the black hole mass and $c$ the speed of light and $r_s$ is the Schwarzschild radius of the solution. Therefore,
\begin{align}
    \phi(r)=\frac{r_s}{r_0}\,\frac{\Gamma(m-\frac{m}{n+2})\Gamma(1+\frac{m}{n+2})}{\Gamma(m)}+O(r^{n+1})\,.
\end{align}
The value of $c_0$,
\begin{align}
    c_0=\frac{r_s}{r_0}\,\frac{\Gamma(m-\frac{m}{n+2})\Gamma(1+\frac{m}{n+2})}{\Gamma(m)}=\frac{r_s}{r_0\,\binom{m-1}{\frac{m}{n+2}}}\,,\label{c0}
\end{align}
should be close to zero for regular solution with asymptotics close enough to Minkowski space, therefore in realistic gravitational theory should be small. This can be achieved by choosing appropriate $m>n>1$ for any given $r_s$ and $r_0$, as seen in \eqref{c0}.

The value of $r_0$ is presumably related not only to the mass of the solution, but also the ``string tension''-like parameter entering the modification of Einstein-Hilbert action, similar to \eqref{r0} and go to zero in the (``infinite tension'') limit, recovering Einstein-Hilbert action and corresponding Schwarzschild solution. Therefore, at least for macroscopic black holes, one can assume $r_s>>r_0$.

\subsection{Exponential suppression around $r=0$ and Minkowski asymptotics}

An interesting example of NED solution can be derived starting from a small modification of \eqref{GCEr},
\begin{align}
    E(r)=\frac{Q\,r^n}{(r^{\frac{n+2}{m}}+\tfrac1m\,r_0^{\frac{n+2}{m}})^m}=\frac{Q}{r^2}\frac{1}{\left(1+\frac{(r_0/r)^{\frac{n+2}{m}}}{m}\right)^m}\,,\label{GCEr1}
\end{align}
in the limit when $n, m\to \infty$ keeping $k=\frac{n+2}{m}\to \frac{n}{m}$ fixed:
\begin{align}
    E(r)=\frac{Q}{r^2}\,e^{-(\frac{r_0}{r})^k}\,,
\end{align}
which reproduces well Coulomb at large distances $r>>r_0$ and exponentially suppresses the $E(r)$ at $r\to 0$.
The potential in the $k=1$ case is given by (up to an additive constant):
\begin{align}
    \phi(r)=\frac{Q}{r_0}\, e^{-\frac{r_0}{r}}\,.
\end{align}
The exponent in this expression provides for suppression at $r\to 0$ that will kill any negative power of $r$. The corresponding double copy metric \eqref{KS} will not have singularities of any curvature invariant, but the asymptotics of the metric will not be Minkowski.

An alternative choice of $\phi(r)$ with the same exponential suppression at $r=0$, which also goes to zero at infinity, is given by
\begin{align}
    \phi(r)=-\frac{Q}{r}\,e^{-(\frac{r_0}{r})^k}\,,
\end{align}
with corresponding electric field:
\begin{align}
    E(r)=\frac{Q}{r^2}\left(1-k\,\left(\frac{r_0}{r}\right)^k\right)\,e^{-(\frac{r_0}{r})^k}\,,\label{ExpE1}
\end{align}
which deviates from the Coulomb at large distances by polynomial term in $\frac{r_0}{r}$, similarly to other cases studied in the previous sections. The behaviour near $r=0$ is not changed though: all poles in curvature invariants are suppressed by the exponential factor. The corresponding metric will be given by \eqref{KS} with:
\begin{align}
    \phi(r)=-\frac{r_s}{r}\,e^{-(\frac{r_0}{r})^k}\,.
\end{align}
This ansatz with exponential suppression at $r=0$ has another advantage: it can provide non-singular solutions for any higher-copy construction of spherically symmetric ``black hole" solutions for higher-spin fields. We do not attempt to answer any questions related to non-linear higher-spin theories, but only note that the generalized Kerr-Schild ansatz is possible for any higher-spin massless field \cite{Didenko:2008va,Didenko:2009td}:
\begin{align}
    h_{\mu_1\mu_2\dots\mu_s}=\phi(r)\,k_{\mu_1}\,k_{\mu_2}\,\dots\,k_{\mu_s}\,,\label{HSKS}
\end{align}
and is also a solution of the free field equations (similarly to Schwarzschild metric that is also a solution for linearized gravity). On the other hand, the multiple-copy interaction picture seems natural for higher-spins, given that their interactions contain building blocks that are higher powers of Yang-Mills interactions (see, e.g., \cite{Fredenhagen:2019lsz} and references therein). Therefore it seems natural to ask the question if the generalisation of this procedure is possible for arbitrary higher spins. Higher-spin potentials involve higher inverse powers of $r$ coming from $k_\mu$'s in \eqref{HSKS} and higher-spin curvatures \cite{deWit:1979sib,Manvelyan:2010jf} contain higher derivatives, therefore, in order to avoid singular curvature invariants, one needs to get rid of higher and higher coefficients in the Taylor expansion of $\phi(r)$ in the corresponding Kerr-Schild ansatz.
Since higher-spin theories contain towers of ever-increasing spin, the non-singular solution requirement would be hard to satisfy unless the function $\phi(r)$ has an exponential suppresion of poles of any order around $r=0$. 
Even though we do not have a strong argument about the necessity of the absence of curvature singularities for higher-spin fields, we note that this condition rules out many candidates for NED-induced double copy black holes and also leads to Minkowski asymptotics.

Taking into account the equation
\begin{align}
    \varphi(r)=\int_r^{\infty} E(x)\,dx\,,
\end{align}
we note that condition for Minkowski asymptotics $\varphi(0)=0$ is satisfied only if $E(r)$ changes the sign somewhere.
Indeed, this is the case for \eqref{ExpE1} and we can deduce an interesting conclusion on the physics of the NED that it solves. For $r<\frac{r_0}{k^{1/{k}}}$ the sign of the electric field changes, therefore, the repelling force of two same-sign charges turns into attraction in short enough distances (and there is an unstable Lagrange point at $r=\frac{r_0}{k^{1/{k}}}$ where $E(r)=0$). Similar phenomenon was observed in example, studied in \cite{Easson:2020esh}. 
If this theory was realistic, there could exist objects with large charge that could grow by swallowing same-sign charges whose energies are high enough to pass the ``Coulomb barrier'' (note that $r_0$ is typically growing with the charge, so the size of the object would be related to the charge: in the explicit examples above, we had $r_0\sim \sqrt{Q}$, or the charge is proportional to the area of the sphere with radius $r_0$).

Even though we cannot derive analytic expression for the Lagrangian, the NED with a solution \eqref{ExpE1} for $k=2$ has a function $\beta(E)$, implicitly given by:
\begin{align}
    E(\beta)=\frac{1-2\, W\left(\frac{\sqrt{e}}{2\, \beta}\right)}{2\, \beta}\,,\label{Ebe}
\end{align}
where $e$ is Euler's constant and $W(z)$ is the real solution of the equation:
\begin{align}
    W\,e^W=z\,.
\end{align}
Given $\beta(s)$, one can reconstruct a Lagrangian of the corresponding NED with $SO(2)$ duality symmetry with the techniques provided in Section \ref{DScompletion}.

\section{Conclusions}\label{Conclusion}

We have studied through specific examples several aspects of Nonlinear Electrodynamics and several features that can pop up in specific examples but are not a general property of arbitrary NED. We discuss below the main results and their implications.

We established that {\it the condition of $SO(2)$ duality symmetry in NED theories is powerful enough to fix the Lagrangian of the theory from the data of the spherically symmetric electrostatic solution}. We interpret this finding in the following crude manner. For NED theories without $SO(2)$ duality symmetry, one would presumably need information from infinite number of solutions with all possible ratios of electric and magnetic charges which is a continuous parameter in this classical context. Thus one would need a function of two variables --- charge ratio and radial coordinate --- to be able to reconstruct the Lagrangian \eqref{L} which is a function of two variables. Instead, the $SO(2)$ duality indicates that all of the solutions with different charge ratios are equivalent, therefore knowing only the electrostatic solution (function of one variable -- radial coordinate) is enough to reconstruct the underlying $SO(2)$-invariant theory (parameterized by a function of one variable \cite{Avetisyan:2021heg}).

If this possibility of reconstructing the theory from solutions generalizes to other cases (with $S-$duality symmetry), it can be a useful tool in many different field and string theories. In particular, it would be desirable to have a systematic method to find a gravity theory for which a given metric is a solution (a recent discussion on this problem can be found in \cite{Knorr:2022kqp}). In particular, would be interesting to find modifications of Einstein-Hilbert action, accommodating solutions with limiting curvature condition (see, e.g., \cite{Frolov:2021afd} and references therein).
When studying the double copy for electrodynamics, we assumed that a NED solution will lead to a solution of some gravity theory, where a mechanism of reverse-engineering the Lagrangian from a solution would be useful. On the other hand, it is not always given that a local purely gravitational Lagrangian with such a solution exists. One can in principle also consider adding sources (see, e.g., \cite{Carrillo-Gonzalez:2017iyj}) to the gravity side, which significantly increases the possibilities. We have seen that as long as we are interested in Coulomb-like static solution, one can generate any such NED solution by an appropriate choice of NED Lagrangian without matter, while, e.g., the ``square root'' of Vaidya metric \cite{Vaidya:1951zz} may require adding sources also to the NED side.

Examples of $SO(2)$-symmetric NED's given by Lagrangians of the form \eqref{L} are rare due to complicated equation \eqref{DS} the Lagrangian has to satisfy. While it is straightforward to write infinite number of such theories in the democratic approach of \cite{Avetisyan:2021heg,Avetisyan:2022zza}, rewriting them in the conventional language of \eqref{L} requires solving the equation \eqref{DS} and is again complicated\footnote{This is due to the ``conservation of mathematical difficulty'', as commented by Oleg Evnin.} in general. 
We provided here a recipe of deriving $SO(2)$-invariant NED theories (used also in \cite{Svazas:2021ltf}) to derive new examples of duality-symmetric NED theories of the form \eqref{L}. We also established a procedure that allows to explicitly reconstruct the $SO(2)$-symmetric Lagrangian from the electrostatic solution.

The constructions provided here can be cumbersome or analytically impossible for some cases. For example, they may require solving a higher-order algebraic equation at some step, for which analytic solutions are not available or do not exist at all. Therefore, we highlight the existence results rather than the concrete construction recipes. 
We expect that for the $SO(2)$-invariant theories the democratic formulation \cite{Avetisyan:2021heg,Avetisyan:2022zza} will be a better guide. 
However, our main target was to elaborate on the double copy construction of regular black holes from NED solutions, building on the work \cite{Pasarin:2020qoa}. This construction is known for the single-potential formulation of gauge theory (democratic formulation, if any, is not yet available for non-abelian gauge theory). There, it was realised that in order for the NED theory to give rise to a regular black hole via double copy procedure, the spherically-symmetric electrostatic solution of the NED should have zero value of the electric field in the centre. This condition is satisfied for everywhere regular solutions that do not require sources --- solitons. One can study general classes of NED's admitting solitonic solutions (such as \eqref{GCEr} with $r_0,m,n>0$) and construct double copy regular black holes. But this is not the only possibility.

While studying some examples of NED theories, we stumbled upon another possibility: that the regular solution of some NED's might require sources of non-zero size. This is a specific feature of some NED theories that, to our best knowledge, have not been studied before. The example of such a theory we studied here is extremely simple, given by a Lagrangian \eqref{ArcsinhLag}. The corresponding source for a spherically-symmetric vacuum solution turns out to be a ball of a finite radius, which somehow ``regularizes'' the infinity of both the electric field value in the proximity of the point charge in Maxwell electrodynamics, and also the self-energy of the solution.\footnote{Similarly to the century-old arguments by Born and Infeld \cite{Born:1934gh}, one can argue that by comparing the finite self-energy of the solution to the rest energy (mass) value of the electron, one can find a radius for it. If we take this point of view seriously, the corresponding size of the electron is not too far away from current experimental reach. We have no reason to consider the theory given by \eqref{ArcsinhLag} realistic, therefore such arguments are only for amusing thought experiments.}
We do not know how to describe the matter whose quanta are these balls in this case: whatever that description is (if it exists) it is not a field theory as we know it, which is suited for describing particles as point singularities.\footnote{We note also, that the natural gravitational analogue of the Lagrangian \eqref{ArcsinhLag} would be of the form $\mathcal{L}=T\,\text{arcsinh}(R/T)$
that would modify Einstein-Hilbert with small corrections for large $T$, possibly resolving the singularity of the Schwarzschild black hole by requiring non-zero size sources, while also giving a model of inflation, somewhat similar to that of \cite{Kruglov:2015ara}.} Clearly, the generalization (if any) of field theory to incorporate such ``particles'' is out of the scope of this work. We hope to come back to this question in the future.

\section*{Acknowledgements}

We are grateful to Zhirayr Avetisyan, Oleg Evnin, Karen Hovhannisyan and Arkady Tseytlin for very useful discussions and comments on this work and to Iwo Bia\l ynicki-Birula for correspondence. We are also indebted to Mariana Carrillo-Gonz\'alez and Axel Kleinschmidt for helpful remarks. The work of KM is supported by the European Union's Horizon 2020 research and innovation programme under the Marie Sk\l odowska-Curie grant number 844265.

\appendix

\section{No birefringence condition and its solutions}\label{NoBirefringence}

Here we summarize the condition of no birefringence for NED \cite{Plebanski:1970zz,Boillat:1970gw} and sketch its solution, showing the uniqueness of Born-Infeld theory among the solutions as a NED theory with approximation to Maxwell. We use the results of Bia\l ynicki-Birula \cite{Bialynicki-Birula:1984daz} which were achieved via analysis of linearization of NED equations. The condition of no birefringence is given by the following differential equations on the NED Lagrangian $\cL(s,p)$ (lower indices indicate derivatives with respective variables):
\begin{align}
\begin{split}
    \cL_{s}\,\cL_{sp}-p\,(\cL_{ss}\,\cL_{pp}-\cL_{sp}^2)=0\,,\\
    \cL_s\,(\cL_{ss}-\cL_{pp})-2\,s\,(\cL_{ss}\,\cL_{pp}-\cL_{sp}^2)=0\,.
\end{split}\label{NBC}
\end{align}
It is straightforward to see that the Maxwell Lagrangian $\cL=-\tfrac12 s$ satisfies the equations \eqref{NBC}, as expected. Another solution, $\cL=-s/p$, was found by Plebanski \cite{Plebanski:1970zz}. We also note, that a Lagrangian of the form $\cL(s,p)=f(p)$ solves the equations \eqref{NBC} for any function $f(p)$. The latter Lagrangians, however, describe theories that do not admit linearization and cannot go to Maxwell theory in any limit, therefore are usually discarded when looking for realistic models of NED. From now on we will assume that $\cL_s\neq 0$. We will make an assumption here, that the Lagrangian can be expanded in powers of $p$ as in \eqref{Lp}:
\begin{align}
    \cL(s,p)=\sum_{n=0}^\infty \frac{p^n}{n!}L_n(s)\,,\label{Lp1}
\end{align}
and by imposing the conditions \eqref{NBC} order-by-order, solve for functions $L_n(s)$. This is a simple exercise, which straightforwardly can be carried out. Note, that, for simplicity, instead of the second equation of the \eqref{NBC}, one can use its linear combination with the first one:
\begin{align}
    \cL_{ss}-\cL_{pp}-\frac{2\,s}{p}\cL_{sp}=0\,.
\end{align}
The result is that $L_{2n+1}(s)=0$, and each $L_{2n}(s)$ is constrained by two conditions:
\begin{align}
    \sum_{k=0}^n \binom{n}{k}\left(2\,s\,L_k^{\prime\prime}\,L_{n-k+2}-2\,s\,L_{k+1}^{\prime}\,L_{n-k+1}^\prime-L_k^\prime\,L_{n-k}^\prime+L_k^\prime\,L_{n-k+2}\right)=0\,,\\
    L_{n+2}+\frac{2\,s}{n+1}L_{n+2}^\prime=L_n^{\prime\prime}\,.
\end{align}
It is not clear why there should exist a solution at all, since the system is overdetermined. It is, however, straightforward to show, that if there is a solution, it is unique (our minimal assumptions are the expansion in the powers of $p$ \eqref{Lp1} and $\cL_s\neq const$).

E.g., for $n=2$ order we get conditions on $L_2(s)$ and $L_0(s)$:
\begin{align}
    L_2+2\,s\,L_2^\prime =L_0^{\prime\prime}\,,\\
    (L_0^\prime+2\,s\,L_0^{\prime\prime})\,L_2=L_0^\prime\,L_0^{\prime\prime}\,,
\end{align}
from where one expresses $L_2$ in terms of $L_0$ and its derivatives, while also get a condition on $L_0$ (assuming $L_0^\prime\neq 0$):
\begin{align}
    L_0^\prime\,L_0^{\prime\prime\prime}-3\,(L_0^{\prime\prime})^2=0\,,
\end{align}
with a solution
\begin{align}
    L_0(s)=c_1\,\sqrt{s+c_2}+c_3\,,
\end{align}
where $c_i$ with $i=1,2,3$ are constants. The further equations will fully determine all $L_{2n}$ in terms of $L_0$ and its derivatives and give more conditions for $L_0$, which may at most fix the values of the constants $c_i$, if there is a solution. We note, however, that the equations contain only derivatives of $L_0$, therefore the $c_3$ cannot be fixed, while all of the equations are homogeneous in powers of different $L_n$'s and therefore cannot fix $c_1$ either. Overall factor and additive constant are not essential in the Lagrangian and cannot be fixed by the system \eqref{NBC}, but can be fixed from the condition that $\cL(s,p)$ can be approximated by Maxwell Lagrangian in some limit. The rest of the equations do not fix any of the $c_i$ as we know that there is still one-parameter freedom in the Born-Infeld theory, which is encoded in the remaining $c_2$. Indeed, the $L_0(s)$ given above is the first term of the Born-Infeld expansion in $p$. Here we will not complete the proof that Born-Infeld theory solves \eqref{NBC}, as it is known. However, we managed to show that there are no other solutions, for which the expansion \eqref{Lp1} holds and $\cL_s\neq const$, confirming the uniqueness of Born-Infeld theory as a physically viable NED without vacuum birefringence, that is, solving \eqref{NBC} (apart from Maxwell).

\section{Gravitational stress-energy tensor and energy for NED solutions}\label{Energy}
The gravitational stress-energy tensor $T^{\mu}_{\;\nu}$  for $d=4$ NED Lagrangians $\mathcal{L} = \mathcal{L}(s,p)$ is given by
\begin{align}
    T^{\mu}_{\;\nu} = - 2\,\frac{\partial \mathcal{L}}{\partial s}F^{\mu\rho}F_{\nu\rho} - 2\frac{\partial \mathcal{L}}{\partial p}F^{\mu\rho}\,{\star}F_{\nu\rho} + \delta^{\mu}_{\;\nu}\mathcal{L}(s,p)\,,
\end{align}
or, using 
$$F^{\mu\rho}\,{\star}F_{\nu\rho}=\frac14\delta^\mu_\nu\, F^{\rho\sigma}\,{\star}F_{\rho\sigma}=\delta^\mu_\nu\,p/2\,,$$
we can rewrite it as
\begin{align}
    T^\mu_\nu=-2\,\frac{\partial\mathcal{L}}{\partial s}(F^{\mu\rho}F_{\nu\rho}-\frac14\,\delta^\mu_\nu\,F^{\rho\sigma}F_{\rho\sigma})-\delta^\mu_\nu\,\left(s\,\frac{\partial\mathcal{L}}{\partial s}+p\,\frac{\partial\mathcal{L}}{\partial p}-\mathcal{L}\right)\,.
\end{align}
The first factor is proportional to Maxwell Stress-Energy tensor and is traceless, while the second factor vanishes for the conformal theories. 

The energy density is given by:
\begin{align}
\label{general energy}
    \mathcal{H} = T^{0}_{0}= - 2\,\beta(s,p)F^{0\rho}F_{0\rho} -p\,\alpha(s,p) + \mathcal{L}(s,p)\nonumber\\
    = 2\,\beta(s,p)E^2 -p\,\alpha(s,p) + \mathcal{L}(s,p)\,.
\end{align}
For spherically symmetric electrostatic solutions, we have 
\begin{align}
    p=0\,, \qquad s=-E^2\,,
\end{align}
and the energy density (\ref{general energy}) reduces to
\begin{align}
    \mathcal{H} &= - 2\,s\,\beta(s) + \mathcal{L}(s)\,,
\end{align}
where $\mathcal{L}(s)=\mathcal{L}(s,p)|_{p=0}$ and $\beta(s)=\beta(s,p)|_{p=0}$.

In addition to that, electrostatic solutions satisfy $-2\beta(s)E(r) = Q/r^2$, which simplifies the energy density further to
\begin{align}
    \mathcal{H} = -\frac{QE(r)}{r^2} + \mathcal{L}(s)\,.
\end{align}
Then, we can compute the self energy of the system by integrating it over the whole space, with the energy density depending only on the radius (we assume background Minkowski spacetime):
\begin{align}
\label{energy}
    H &= \int_{\mathbb{R}^3}\text{d}\textbf{x}^3\;\mathcal{H} =4\pi\int_{\mathbb{R}^+}\text{d}r\;r^2\,\mathcal{H} \nonumber\\[2pt]
    &=-4\pi Q\int_{\mathbb{R}^+}\text{d}r\;E(r) + 4\pi\int_{\mathbb{R}^+}\text{d}r\;r^2\,\mathcal{L}(-E^2)
\end{align}
where we used that in the elctrostatic solution $s = -E^2(r)$. Therefore, we only need to know the electrostatic solutions and the Lagrangian to get the energy of the electrostatic theories.
We label the terms of the energy as follows:
\begin{align}
\label{notation}
    H &\equiv H_1 + H_2\,,\\[4pt]
    H_1 &\equiv -4\pi Q\int_{0}^{\infty}\text{d}r\;E(r)\,, \\[5pt] 
    H_2 &\equiv 4\pi\int_{0}^{\infty}\text{d}r\;r^2\mathcal{L}(-E^2(r))\,.
\end{align}
We note also, that in the electrostatic case $E(r)=\frac{\partial \varphi(r)}{\partial r}$, where $\varphi(r)=A_0(r)$ is the electrostatic potential. Then,
\begin{align}
    \int_{0}^\infty d r\,E(r)=\varphi(r)|_{r=\infty}-\varphi(r)|_{r=0}=-\varphi(0)\,,
\end{align}
where in the last equality we assumed that the electrostatic potential vanishes at the infinity. Therefore,
\begin{align}
    H_1=4\pi\,Q\,\varphi(0)\,.
\end{align}
For solutions that have $\varphi(0)=0$, corresponding to Minkowski asymptotics of the double copy black hole, we get $H_1=0$ and $H=H_2$.

When trying to compute the Energy of the solution \eqref{Ex1}, one needs to take into account the addition to the Lagrangian accounting for the coupling to the matter that makes up the ``particle''. This can be done by adding a term $A_\mu\,j^\mu$ where the $j^\mu=(\rho(r),0,0,0)$, treating it as a static matter. 

\end{document}